\documentclass[aps,prl,nofootinbib,twocolumn,superscriptaddress,floatfix,notitlepage,10pt]{revtex4-2}
\usepackage[export]{adjustbox} 
\usepackage[utf8]{inputenc}
\usepackage[dvipsnames]{xcolor}
\usepackage{graphicx}
\usepackage{amsmath,amsfonts,amssymb}
\usepackage[breaklinks,colorlinks,urlcolor=Plum,citecolor=MidnightBlue,linkcolor=WildStrawberry]{hyperref}

\usepackage{bm}
\usepackage{acronym}
\usepackage{xspace}
\usepackage{multirow}
\usepackage{tabularx}
\usepackage{orcidlink}
\usepackage{ragged2e}
\usepackage{setspace}
\usepackage{caption}
\usepackage{subcaption}
\usepackage{comment}
\newcommand{\msun}{M_{\odot}}

\newcommand{\td}{{\rm d}}

\newcommand{\dptidal}{\delta\psi_{\mathrm{tidal}}}

\begin{document}

\title{No Love for black holes: tightest constraints on tidal Love numbers of black holes from GW250114}
\author{M. Andr\'es-Carcasona\orcidlink{0000-0002-8738-1672}}
\email{mandresc@mit.edu}
\affiliation{LIGO Laboratory, Massachusetts Institute of Technology, Cambridge, MA 02139, USA}
\affiliation{Kavli institute for Astrophysics and Space research, Massachusetts Institute of Technology, Cambridge, MA 02139, USA}
\author{G. Caneva Santoro\orcidlink{0000-0002-2935-1600}}%
\email{giada.santoro@nbi.ku.dk}
\affiliation{Center of Gravity, Niels Bohr Institute, Blegdamsvej 17, 2100 Copenhagen, Denmark}%

\date{\today}

\begin{abstract}
Tidal Love numbers of black holes, zero in classical general relativity for Kerr black holes in vacuum, become non-vanishing in the presence of exotic matter or in alternative theories of gravity, making them a powerful probe of fundamental physics. The gravitational-wave event GW250114, observed with an unprecedented signal-to-noise ratio, provides a unique opportunity to test this prediction. By analyzing this event, we conclude that the data is consistent with the binary black hole hypothesis, and we place a 90\% upper limit on the effective tidal deformability of $\tilde{\Lambda} < 34.8$. These bounds imply that any environment surrounding the black holes must contribute less than $\sim  7\times 10^{-3}$ of their mass, and they rule out some models of boson stars. Our findings provide the strongest observational constraints yet on black hole tidal deformability and show that the data remain fully consistent with the Kerr black hole prediction of vanishing tidal Love numbers.
\end{abstract}
\maketitle

\textbf{\textit{Introduction}} -- The first direct detection of gravitational waves (GWs) from a binary black hole (BH) in 2015~\cite{LIGOScientific:2016aoc} opened a new era of GW astronomy, enabling precision tests of general relativity (GR) in the strong-field regime and probing the nature of compact objects. Since then, the LIGO–Virgo–KAGRA (LVK) collaboration has expanded the catalog to GWTC-4, containing hundreds of compact-binary coalescences~\cite{LIGOScientific:2018mvr,LIGOScientific:2020ibl,KAGRA:2021vkt,LIGOScientific:2025slb,LIGOScientific:2025hdt,LIGOScientific:2025yae}. To date, all observations are consistent with the predictions of GR, with no statistically significant deviations reported~\cite{LIGOScientific:2016lio,LIGOScientific:2025jau,LIGOScientific:2025cmm,LIGOScientific:2021sio}.
Within the post-Newtonian (PN) description of the inspiral, tidal interactions appear as high-order corrections to the GW phase. Their strength is quantified by the tidal Love numbers (TLNs)~\cite{Flanagan:2007ix,Vines:2011ud,Blanchet:2013haa}, which measure the multipolar response of a body to an external tidal field. In vacuum GR, BHs have vanishing TLNs~\cite{Binnington:2009bb,Ivanov:2022qqt,Cardoso:2017cfl,Berti:2015itd,Chia:2020yla,Charalambous:2021mea,LeTiec:2020bos,DeLuca:2023mio}, reflecting the fact that their horizons cannot sustain static tidal deformations. A robust measurement of nonzero TLNs would therefore indicate physics beyond vacuum GR, such as modified gravity, exotic compact objects (ECOs), or the presence of matter surrounding the binary~\cite{DeLuca:2021ite,Cardoso:2018ptl,Pani:2015tga,Mayerson:2020tpn,Cardoso:2019vof,Consoli:2022eey,Saketh:2023bul,Cardoso:2019upw,DeLuca:2022tkm}. Environmental effects provide one possible mechanism for inducing non-zero TLNs. For example, accretion disks, boson clouds, or dark-matter halos can imprint tidal signatures on the waveform~\cite{Baumann:2018vus,DeLuca:2021ite,Brito:2023pyl,Capuano:2024qhv,Cardoso:2019upw,Cardoso:2021wlq,Katagiri:2023yzm,DeLuca:2024uju,Roy:2025qaa}. These effects grow stronger for denser environments and may also appear at other PN orders or during the ringdown phase~\cite{Barausse:2014pra,Barausse:2014tra,Alnasheet:2025mtr,CanevaSantoro:2023aol,Lan:2025brn,Alnasheet:2025mtr,Pezzella:2024tkf,Berti:2025hly,Leong:2023nuk}. Neglecting such imprints can bias the recovered astrophysical parameters~\cite{CanevaSantoro:2023aol,DeLuca:2025bph}.
Previous searches for TLNs in BBH mergers using LVK data yielded only weak constraints, due to the limited signal strength and the fact that TLNs enter the GW phase at high PN order. The first dedicated studies from Refs.~\cite{Narikawa:2021pak,Chia:2023tle} found no evidence for nonzero TLNs.

On 14 January 2025, during the second part of the O4 observing run (O4b), a signal was observed in the LIGO Hanford and LIGO Livingston interferometers and designated as GW250114~\cite{KAGRA:2025oiz,LIGOScientific:2025obp}. This event shares several properties with GW150914 but was detected with a significantly higher signal-to-noise ratio (SNR), of approximately 80, thanks to an improved detector sensitivity~\cite{Capote:2024rmo,LIGO:2024kkz}. The enhanced SNR allows a more detailed characterization of the source. Indeed, GW250114 has already been used to confirm Hawking’s area theorem~\cite{KAGRA:2025oiz} and to place the strongest constraints on GR deviations from a single GW event to date~\cite{LIGOScientific:2025obp}.
The high SNR is particularly beneficial for probing high-PN corrections in the inspiral, where tidal effects enter. While Ref.~\cite{LIGOScientific:2025obp} performed PN deviation tests, their analysis extended only to 3.5PN and, therefore, could not access TLN contributions. In this \textit{Letter}, we use data from GW250114 to place the most stringent upper limits to date on the TLNs of BHs.
Throughout this work we use geometrical units ($G=c=1$).

\textbf{\textit{Methods}} -- In the adiabatic limit, an external tidal field $\mathcal{E}_{ij}$ induces a quadrupole moment $Q_{ij}$ in a compact object given by~\cite{Flanagan:2007ix,Hinderer:2007mb,Binnington:2009bb,Hinderer:2009ca}
\begin{equation}\label{eq:Qij}
    Q_{ij} = -\Lambda m^5 \mathcal{E}_{ij}, \qquad \Lambda = \frac{2}{3} k_2 ,
\end{equation}
where $\Lambda$ is the dimensionless tidal deformability, $m$ is the mass of the object, and $k_2$ the quadrupolar ($l=2$) tidal Love number. We adopt the convention of Ref.~\cite{Cardoso:2017cfl}, where the object's compactness is absorbed into the definition of $k_2$, in contrast to other conventions where it appears explicitly. The value of $k_2$ depends on the internal structure of the object. For neutron stars, $k_2 \sim 210$, although its exact value depends on the equation of state~\cite{Binnington:2009bb,Damour:2009vw,Yagi:2013awa,Amin:2022pzv,Riley:2019yda,Hinderer:2007mb}. For exotic compact objects such as boson stars or wormholes $k_2$ can span several orders of magnitude~\cite{Cardoso:2017cfl,Ryan:2022hku,Brito:2015pxa,Herdeiro:2020kba,Diedrichs:2023trk,Jain:2021pnk}.

Tidal deformabilities imprint a distinct effect on the gravitational-wave phase, entering at high post-Newtonian order, starting at 5PN and 6PN, where they contribute an additional phase term~\cite{Flanagan:2007ix,Vines:2011ud}
\begin{multline}
\label{eq:psi_tidal}
        \dptidal = -\frac{3}{128 \eta }\left[ \frac{32}{9}\tilde{\Lambda} \, v^{5} + \frac{3115}{64}\tilde{\Lambda} \, v^{6} \right .
        - \\ 
        \left. \frac{6595}{364}\delta\tilde{\Lambda}\sqrt{1 - 4\eta}\, v^{6} \right] ,
\end{multline}
where $v=(\pi M f)^{1/3}$, $f$ is the GW frequency, $M=m_1+m_2$ the total mass, and $\eta=m_1 m_2 / M^2$ the symmetric mass ratio,
\begin{multline}
\tilde{\Lambda} = \frac{8}{13} \Big[ (1 + 7\eta - 31\eta^2)(\Lambda_1 + \Lambda_2) 
\\+ \sqrt{1 - 4\eta}\,(1 + 9\eta - 11\eta^2)(\Lambda_1 - \Lambda_2) \Big] ,
\end{multline}
and
\begin{multline}
\delta \tilde{\Lambda} = \frac{1}{2} \Bigg[
\sqrt{1 - 4\eta}\left(1 - \frac{13272}{1319}\eta + \frac{8944}{1319}\eta^2 \right)(\Lambda_1 + \Lambda_2)
\\+
\left(1 - \frac{15910}{1319}\eta + \frac{32850}{1319}\eta^2 + \frac{3380}{1319}\eta^3 \right)(\Lambda_1 - \Lambda_2)
\Bigg] .
\end{multline}

These expressions account only for the \textit{electric} TLNs associated with the mass quadrupole in Eq.~\eqref{eq:Qij}. We neglect the \textit{magnetic} TLNs, which correspond to current multipoles~\cite{Abdelsalhin:2018reg,Cardoso:2017cfl}, as well as any higher-order \textit{electric} multipoles.

At the waveform level, we treat $(\Lambda_1,\Lambda_2)$ as additional intrinsic parameters. We incorporate the tidal phase correction into the phenomenological inspiral–merger–ringdown waveform \texttt{IMRPhenomPv2}~\cite{Husa:2015iqa,Khan:2015jqa,Hannam:2013oca}. The waveform is modified by adding two generic absolute dephasing corrections to the inspiral phase,
\begin{equation}
    \psi^{\rm tidal}(f) = \psi^{\rm vac}(f) + \sum_{j\in\{10,12\}} \frac{3}{128 \eta } \, \delta\psi_j \, v^{\,j-5}\, .
\end{equation}

The coefficients $\delta\psi_j$ are obtained by matching this parameterized form to Eq.~\eqref{eq:psi_tidal}. This approach follows the model-agnostic framework of parameterized tests of GR~\cite{Agathos:2013upa,Roy:2025gzv,Li:2011cg,Meidam:2017dgf}, widely used to search for deviations at specific PN orders~\cite{LIGOScientific:2016lio,LIGOScientific:2021sio,DeLaurentis:2016jfs,CanevaSantoro:2023aol,LIGOScientific:2018dkp,LIGOScientific:2019fpa,LIGOScientific:2020tif}. and we implement this dephasing by modifying the standard \texttt{LALSuite} library~\cite{Wette:2020air,lalsuite}. Although exotic physics or environmental effects could imprint deviations also in the post-inspiral regime, previous analysis have shown that the remnant is consistent with a Kerr BH~\cite{KAGRA:2025oiz,LIGOScientific:2025obp}. We therefore focus our search on the inspiral phase, where tidal effects are well-modeled within the PN framework, and leave the merger–ringdown portion of the waveform unchanged.

We analyze the cleaned, calibrated strain data from LIGO Hanford and Livingston, centered on the GPS trigger time $t_0 = 1420878141$~s and processed following Ref.~\cite{KAGRA:2025oiz}. For each detector, we select a segment spanning 6~s before to 2~s after the trigger, resample to 2048~Hz, restrict the analysis to $20$–$896$~Hz, and apply a Tukey window with a 1~s roll-off.

We use a Bayesian framework to achieve two goals. First, to test the null hypothesis of vanishing Love numbers ($\mathcal{H}_{0}: \Lambda_i=0$) against the alternative that they are non-zero ($\mathcal{H}_{1}: \Lambda_i\neq0$) by computing the Bayes factor, $\mathcal{B}^{\Lambda_i\neq0}_{\Lambda_i=0}$. Second, to perform parameter estimation under the $\mathcal{H}_{1}$ hypothesis to derive constraints on the tidal deformability parameters. We compute the posterior distributions and the Bayes factor using the \texttt{bilby} library~\cite{Ashton:2018jfp,Romero-Shaw:2020owr} with the \texttt{dynesty} nested sampler~\cite{Speagle:2019ivv}.

We adopt standard priors for the binary intrinsic and extrinsic parameters: uniform in detector-frame masses and spin magnitudes; isotropic in spin and orbital orientations; and uniform in comoving volume with isotropic sky position. For the tidal deformabilities $\Lambda_i$, we employ log-uniform priors on the range $[10^{-3},5000]$. This scale-invariant choice avoids the prior-volume preference for large $\Lambda_i$ inherent in uniform priors when the data are only weakly informative at high PN order. Results obtained using uniform priors, following Ref.~\cite{Chia:2023tle}, are provided in the the Supplemental Material for completeness. We also marginalize over calibration uncertainties using a cubic-spline model with 10 amplitude and 10 phase control points~\cite{CalUncer}. The full prior table is given in the Supplemental Material.

\textbf{\textit{Results}} -- Our analysis of GW250114 reveals no evidence for tidal deformability, allowing us to place a stringent 90\% upper limit on the effective tidal deformability of $\tilde{\Lambda} < 34.8$. This null result is corroborated by the log-Bayes factor comparing the tidal and non-tidal hypotheses, which we compute to be $\ln\mathcal{B}^{\Lambda_i\neq0}_{\Lambda_i = 0}=-0.063\pm 0.182$. A value this close to zero indicates that the data has no statistical preference for the more complex tidal model over the simpler BBH hypothesis, in line with the predictions of general relativity for Kerr BHs in vacuum. We release the data produced in~\cite{DataRelease_zenodo}.

\begin{figure*}[htbp]
    \centering
    \includegraphics[width=0.5\textwidth]{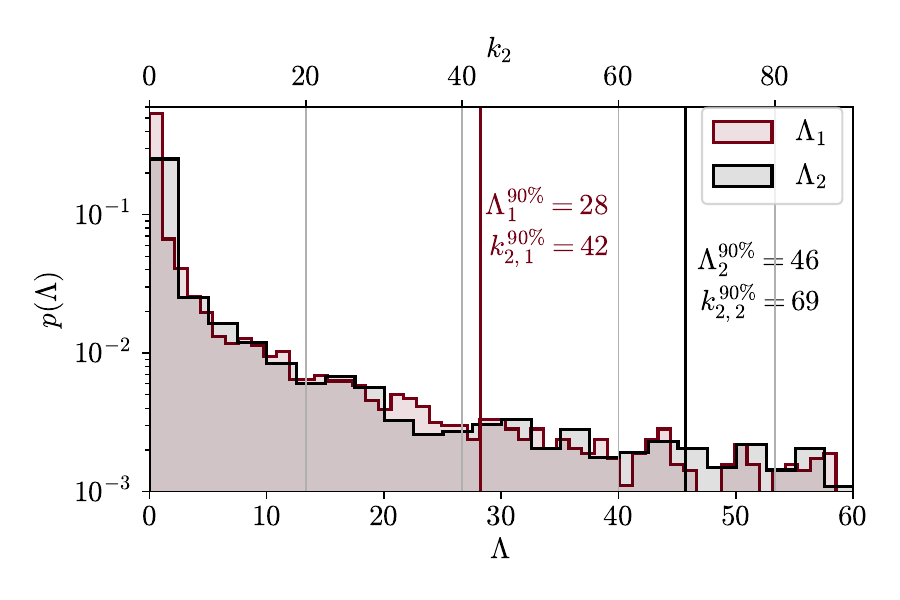}
    \includegraphics[width=0.44\textwidth]{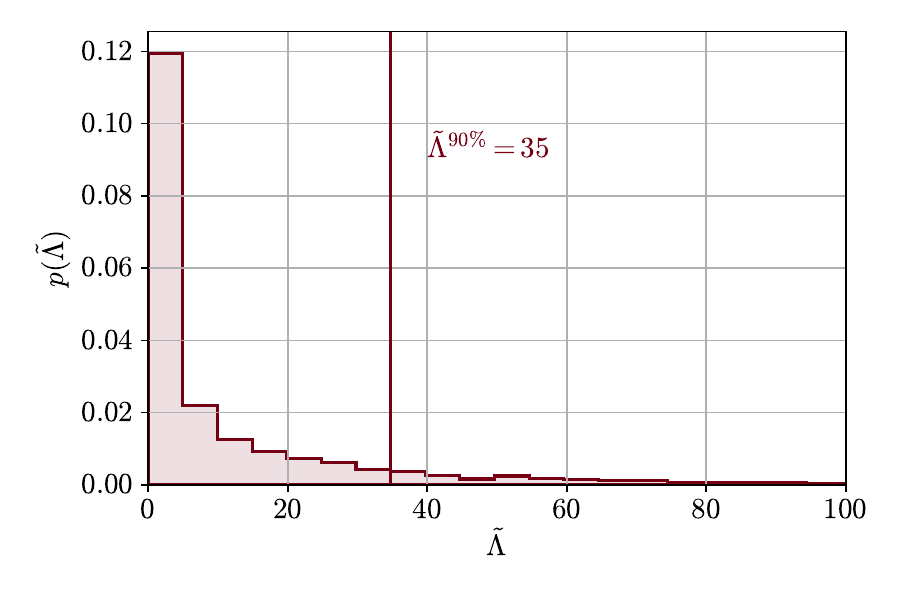}
    \caption{\textit{(Left}) Marginalized posterior distributions for individual tidal deformabilities $\Lambda_i$ and tidal Love numbers $k_2$ for GW250114. Vertical lines indicate 90\% credible upper limits. (\textit{Right}) Marginalized posterior distribution for the effective tidal deformability $\tilde{\Lambda}$. The vertical line shows the 90\% credible upper bound.}
    \label{fig:LambdaLimits}
\end{figure*}

Figure~\ref{fig:LambdaLimits} presents the posterior distributions for the main tidal parameters. The right panel shows the posterior for the effective tidal deformability $\tilde{\Lambda}$, the parameter to which the waveform is most sensitive. This is because $\tilde{\Lambda}$ enters at the leading (5PN) tidal order, and the next-order tidal term, $\delta \tilde{\Lambda}$, is suppressed for a nearly symmetric binary like GW250114. Our posterior strongly peaks near zero, confirming the null result. The left panel shows the marginalized posteriors for the individual component deformabilities, $\Lambda_1$ and $\Lambda_2$, which also show clear support for vanishing values. In the same panel we also show the values of the individual TLNs.

From these distributions, we derive 90\% credible upper limits of $\Lambda_1 < 28.2$ and $\Lambda_2 < 45.7$. This constraint on $\Lambda_i$ represents an improvement of more than an order of magnitude over the tightest limits from previous GW catalogs~\cite{Narikawa:2021pak,Chia:2023tle}. The values for the individual tidal deformabilities are obtained by inverting Eq.~\eqref{eq:Qij} which translates these into the 90\% upper limits on the individual TLNs of $k_{2,1} < 42.4$ and $k_{2,2} < 69.5$. These are the most stringent observational constraints on BH tidal Love numbers, and follow from the extremely high SNR with which GW250114 has been observed.

\begin{figure}[htbp]
    \centering
    \includegraphics[width=\columnwidth]{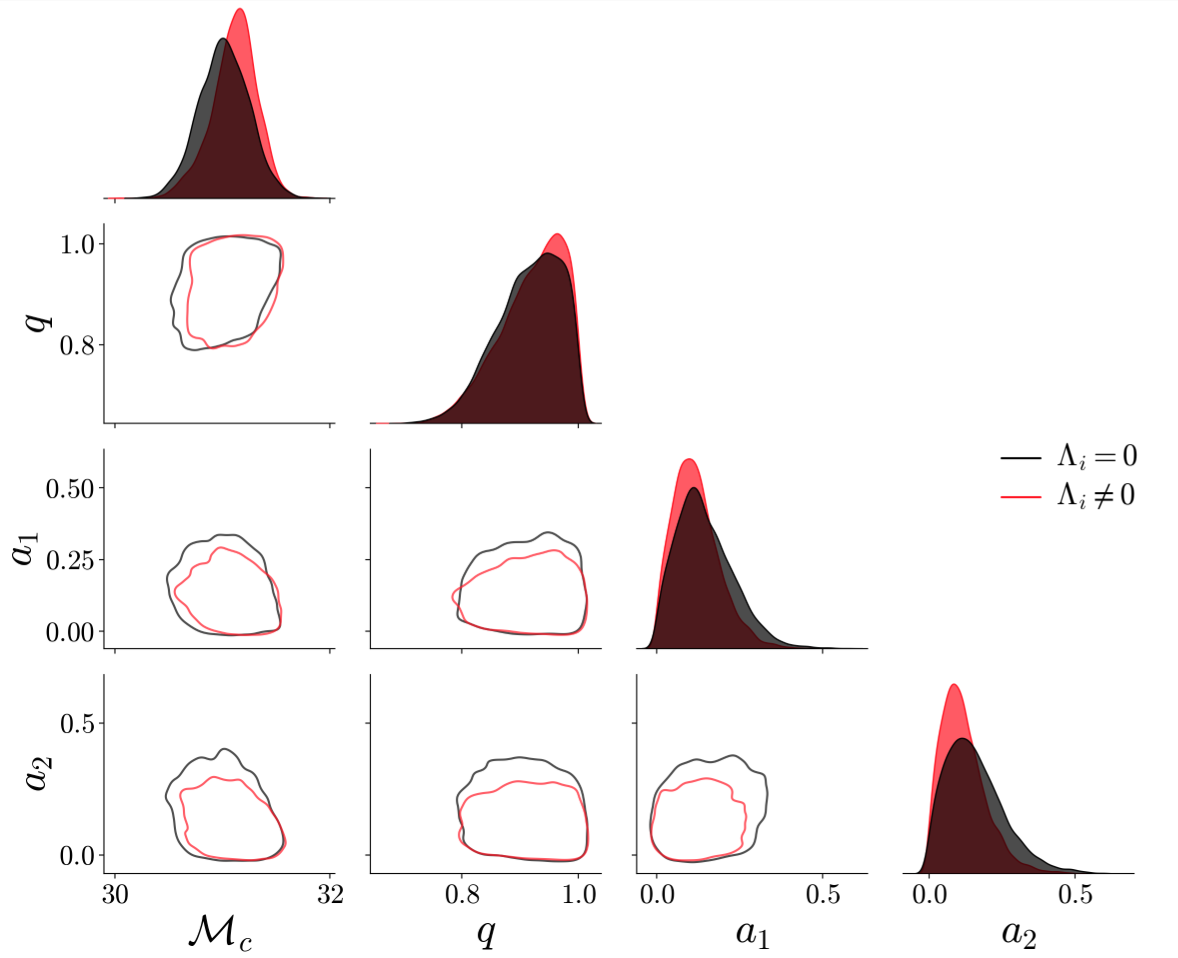}
    \caption{Corner plot comparing the posterior distributions for the chirp mass, mass ratio and spins with (red) and without (black) TLNs.}
    \label{fig:logU_corner}
\end{figure}

\begin{figure}[htbp]
    \centering
    \includegraphics[width=\columnwidth]{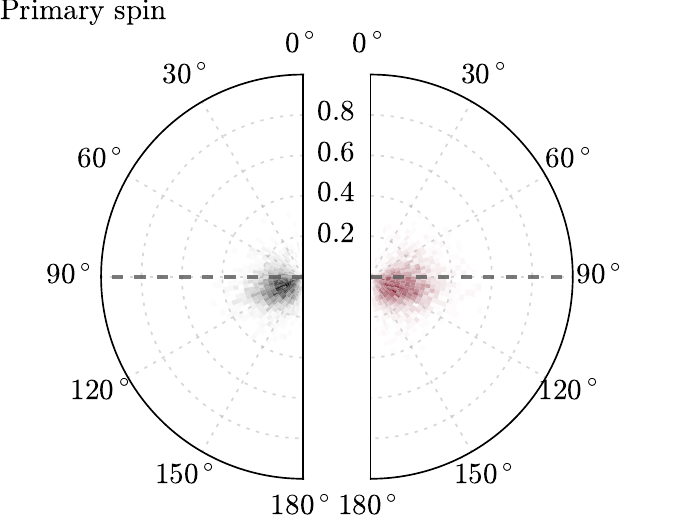}
    \includegraphics[width=\columnwidth]{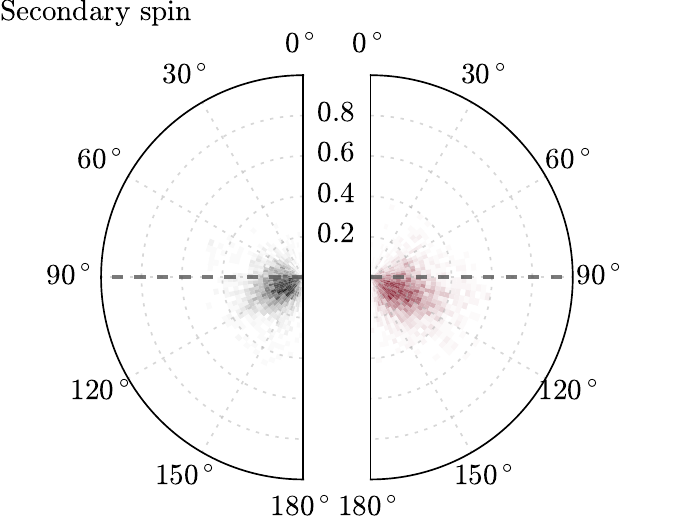}
    \caption{Comparison of posterior distributions for the orientation and magnitude of the primary (top) and secondary (bottom) spin components, for models with $\Lambda_i = 0$ (black) and $\Lambda_i \neq 0$ (red). A spin angle of zero corresponds to perfect alignment with the orbital angular momentum.}
    \label{fig:spins}
\end{figure}

To assess the impact of tidal parameters on the inference of other source properties, Fig.~\ref{fig:logU_corner} compares posterior distributions for key intrinsic parameters with (red) and without (black) tidal effects. The posteriors for chirp mass $\mathcal{M}_c \equiv (m_1m_2)^{3/5}/(m_1+m_2)^{1/5}$, mass ratio $q \equiv m_2/m_1$, and individual spin magnitudes ($a_1, a_2$; shown in Fig.~\ref{fig:spins}) all show remarkable stability, with excellent agreement between the two analyses. Similarly, the spin orientation posteriors remain unchanged and close to the orbital plane.

This stability is a crucial validation of our analysis as it shows that there is no need to bias other astrophysical parameters to accommodate the potential for tidal effects, even though that there are degeneracies among them. The consistency between the two models confirms that the simpler, non-tidal hypothesis is sufficient to describe the data, and it is important as unaccounted non-vanishing TLNs in the signal would strongly bias the mass and spins estimations~\cite{DeLuca:2025bph}. The minimal nature of these shifts, especially in the mass parameters, is a direct consequence of using a LogUniform prior in the tidal deformabilities, which places a prior support arbitrarily near zero and allows the data to favor a non-tidal solution. This stands in contrast to the effect of a Uniform prior for such high SNR event, which, as we demonstrate explicitly in the Supplemental Material, induces a significant artificial bias in the mass and spin parameters. This bias in the mass posterior was already reported in Ref.~\cite{Chia:2023tle}.

These results show a self-consistent conclusion. The negligible Bayes factor, the posteriors for all tidal parameters peaking at zero, and the stability of the inferred intrinsic parameters provide multiple, consistent lines of evidence that GW250114 is a binary BH merger whose signal is indistinguishable from the predictions of general relativity in vacuum.

\textbf{\textit{Implications}} --  Our findings carry significant implications for the nature of the compact objects and their potential astrophysical environments. The consistency with zero tidal deformability provides stringent constraints on exotic scenarios beyond the standard BH in vacuum paradigm.

\textit{Constraints on environmental effects}-- If the binary system is embedded in an external environment of mass $m_{\rm env}$,  with $m_{\rm env} \ll m_{\rm BH}$ (i.e., $\varepsilon \equiv m_{\rm env}/m_{\rm BH} \ll 1$) and characteristic scale $L$, the GW would be characterized by non-vanishing TLNs~\cite{DeLuca:2021ite,DeLuca:2024uju,DeLuca:2025bph}. While specific environmental configurations yield different relations between tidal deformability and environmental parameters, a generic scaling of the form $k_2 \propto (L/m_{\rm BH})^5 $ holds~\cite{Baumann:2018vus,DeLuca:2021ite}. Following the approach of Ref.~\cite{DeLuca:2025bph}, we parameterize the effective environmental tidal deformability as \begin{equation} \label{eq:env}
    \tilde{\Lambda} = \frac{2}{3}\mathcal{F}\varepsilon\tilde{L}^5\, ,
\end{equation} 
where $\tilde{L} \equiv L/m_{\rm BH}$ and $\mathcal{F}$ is an $\mathcal{O}(1)$ factor encoding the details of the nature of the actual environment. Setting $\mathcal{F} = 1$ provides conservative constraints that can be rescaled for specific models.

Physical environments require $\tilde{L} \gtrsim 6$ to maintain equilibrium~\cite{DeLuca:2025bph} and the most conservative mass constraints of the environment occur at this minimal scale. Our 90\% credible upper limit on the effective tidal deformability translates to an environmental mass fraction of $\varepsilon^{90\%} = 7\times 10^{-3}$ at $\tilde{L} = 6$. These constraints become progressively tighter for larger environments due to the $\tilde{L}^5$ scaling of Eq.~\eqref{eq:env}.

Nonetheless, we note that we are not modeling the potential tidal disruption of the environment, which would make the tidal deformability change as the binary evolves and eventually vanish once the environment gets disrupted. Therefore, extending these constraints to larger values $\tilde{L}$ without specifically modeling this effect can lead to some systematic biases at the order of $\mathcal{O}(10\%)$, as addressed in Ref.~\cite{DeLuca:2025bph}. We provide in the Supplemental Material more details about the regime of validity of these constraints.

\textit{Constraints on exotic compact objects}-- Our results can also be used to provide stringent constraints on various exotic compact object scenarios. Following the formalism of Ref.~\cite{Cardoso:2017cfl} and its data release~\cite{DataBS} we can test specific exotic compact object models that predict positive, non-zero, TLNs. In particular, we focus on boson stars (BS), but our agnostic results can be recast to any model. We consider three primary BS models, each characterized by different scalar field potentials: minimal, massive, and solitonic configurations (see the Supplemental Material for details).

\begin{figure}[htbp]
    \centering
    \includegraphics[width=1.0\columnwidth]{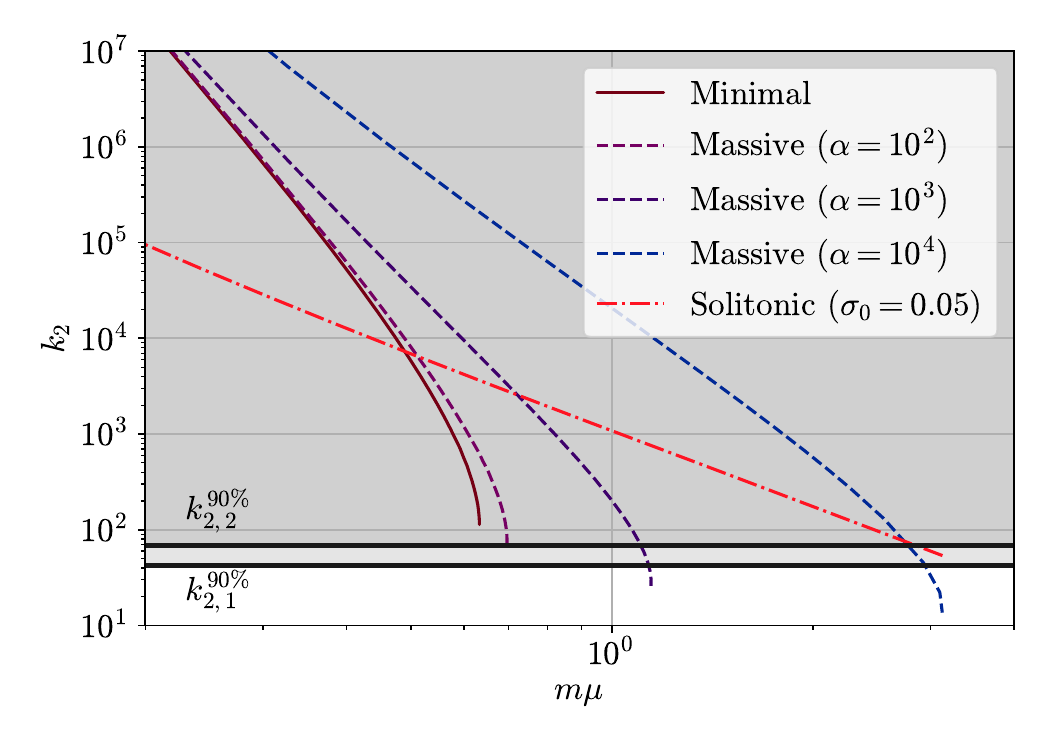}
    \caption{Value of the TLN as a function of the product $m\mu$ for various models reported in Ref.~\cite{Cardoso:2017cfl}. Shaded regions indicate excluded regions by the results of our analysis.}
    \label{fig:BSlimits}
\end{figure}

We show in Fig.~\ref{fig:BSlimits} a plot of the predicted TLNs for the various models as a function of the product $m\mu$, where $m$ is the mass of the BS and $\mu$ the mass of the boson. We include in the figure the limits at the 90\% confidence level that we have obtained on the TLNs, showing the ruled out region as the shaded area.

For minimally coupled boson stars, the stable ground-state sequence reaches a minimum $k_2 \simeq 113.5$ at $m\mu \simeq 0.633$~\cite{Cardoso:2017cfl,DataBS}. Our 90\% upper bounds of $42.4$ and $69.5$ for the primary and secondary components, respectively, excludes a mini-BS interpretation with more than a 90\% confidence level and consequently rules out a mini-BS binary system for GW250114.

Within the massive boson star model with coupling strength $\alpha = 100$~\cite{Cardoso:2017cfl}, the tidal deformability decreases monotonically with increasing $m\mu$, reaching $k_2 \simeq 68.3$ near the branch endpoint at $m\mu \simeq 0.696$. Our constraints imply that the primary component cannot be a BS, while for the secondary only the very high compactness region remains viable at the 90\% confidence level. Nonetheless, as the primary object cannot be a BS under this model, we can also rule out the BS-BS merger hypothesis if the coupling strength is $\alpha=100$. However, increasing the coupling strength to $\alpha = 10^3$ ($\alpha = 10^4$) relaxes these constraints a bit. The allowed regions at the 90\% confidence level are $m_1\mu\in [1.135,1.144]$ ($m_1\mu\in [2.950,3.124]$) and $m_2\mu\in [1.1038,1.144]$ ($m_2\mu\in [2.822,3.124]$). These values indicate that while the BS-BS under these strengths cannot be completely ruled out, the viable window is pretty much confined to the high-compactness configurations. In fact, using also the posterior samples for the masses, we can recast it to limits to the boson mass itself also considering that both BS are formed by the same bosonic field. The 90\% credible intervals become  $\mu \in [7.3\times 10^{-12}, 8.1\times 10^{-12}]$~eV and $\mu \in [1.8\times 10^{-11}, 2.2\times 10^{-11}]$~eV for $\alpha= 10^3$ and $\alpha = 10^{4}$, respectively.

Finally, for the solitonic boson star model with $\sigma_0 = 0.05$~\cite{Cardoso:2017cfl}, our 90\% confidence level constraints are able to rule out that primary object is a BS of this kind, because the maximum compactness configuration for this case happens at $m\mu = 3.126$ with a $k_2=53.8$~\cite{DataBS}. As a direct consequence, we confidently rule out the BS-BS hypothesis for this solitonic model.

Other ECOs, such as wormholes, perfect mirrors or gravastars, would have negative TLNs~\cite{Cardoso:2017cfl}, implying that they would contract due to the gravitational field created by the binary companion. Since we have restricted our analysis to positive TLNs, we cannot make statements about these other ECOs.

\textbf{\textit{Conclusions}} -- 
We have performed a high-precision test for tidal deformations using the high SNR event GW250114. Our analysis reveals no evidence for tidal effects, allowing us to place the most stringent upper limit on the effective tidal deformability to date: $\tilde{\Lambda} < 34.8$ at 90\% credibility. This null result is quantified by a log-Bayes factor of $\ln\mathcal{B}^{\Lambda_i\neq 0}_{\Lambda_i= 0} \approx 0$, indicating that the data show no statistical preference for a tidal model over the simpler binary BH hypothesis. These results translate into individual component constraints of $\Lambda_1 < 28.2$ and $\Lambda_2 < 45.7$ at the 90\% confidence level, improving upon previous limits by more than an order of magnitude. Our inferred astrophysical parameters, such as masses and spins, show no clear bias when a waveform using tidal effects is employed.

Our upper limits have some implication for fundamental physics and the nature of the environments surrounding this binary. It constrains the mass of any potential astrophysical environment to contribute less than $\sim 0.7\%$ of each BH's mass. Furthermore, our constraints on the individual TLNs allow us to exclude several BS scenarios. We rule out a binary consisting of either minimally coupled BS, massive BS with a coupling of $\alpha=100$, or solitonic boson stars with $\sigma_0=0.05$. For the massive boson star models that remain viable (e.g., with $\alpha \ge 10^3$), our analysis provides tight constraints on the boson mass.

\vspace{5mm}
\begin{acknowledgments}
G.C.S. thanks Vitor Cardoso and Ariadna Ribes Metidieri for insightful comments and discussions during the development of this project. The authors would also like to thank the feedback received by Nicola Franchini, Vasco Gennari and Juan Calderon Bustillo. The Center of Gravity is a Center of Excellence funded by the Danish National Research Foundation under grant no. DNRF184. The authors are grateful for computational resources provided by the LIGO laboratory and supported by National Science Foundation Grants PHY-0757058 and PHY-0823459. This material is based upon work supported by NSF’s LIGO Laboratory which is a major facility fully funded by the National Science Foundation and operates under Cooperative Agreement No. PHY-1764464. 
\end{acknowledgments}

\bibliography{ref}

\appendix
\section{Details of the analysis}\label{sec:PNcorrections}

The PN-phase deviations during the inspiral are implemented in the \texttt{IMRPhenomPv2} waveform of the \texttt{LALSuite}~\cite{lalsuite} package by exploiting its reliance on the \texttt{TaylorF2} approximation in the inspiral regime. During this regime, the strain is approximated, in the frequency domain, as
\begin{equation}
    \tilde{h}(f) = \mathcal{A}f^{-7/6}\exp\left[ i\psi(f) \right]\, ,
\end{equation}
where $\mathcal{A}$ is the amplitude and $\psi(f)$ the phase. Since the tidal corrections add linearly to the point-particle phase of the \texttt{TaylorF2}, the phase simply becomes~\cite{Wade:2014vqa}

\begin{equation}
    \psi(f) = \psi_{\mathrm{pp}}(f)+\dptidal(f)\, .
\end{equation}

In the \texttt{IMRPhenomPv2}, the total phase is built piecewise from inspiral, intermediate and merger-ringdown sectors using smooth window functions to ensure $\mathcal{C}^{1}$ continuity at the transition frequencies.Since our tidal modification is applied at the phase level only during the inspiral, we subtract the constant and linear contributions of the tidal phase at a reference frequency to avoid shifting the coalescence time and phase, which are re-fit as free parameters. This is the same
stitching strategy used in LAL for testing GR with this PN deviation formalism. In our case, we leave the amplitude of the waveform unmodified. 

We note that the waveform that we employ only contains the dominant mode, while other sub-dominant harmonic modes have been observed in GW250114. Nonetheless, their contribution is modest due to the nearly symmetric nature of the binary. For instance, the $(l, |m|) = (4, 4)$ mode is reported with a network SNR of $\sim 3.6$~\cite{LIGOScientific:2025obp}. Although small, this unmodeled power could introduce a systematic bias in the posterior, as degeneracies linking tidal parameters and higher-order modes through the mass ratio may cause the tidal parameters to partially absorb this residual signal. For a null test such as the one we carry, any such effect would tend to push the posterior for the tidal deformability away from zero. We therefore conclude that the upper limit we present on $\tilde{\Lambda}$ should be regarded as a conservative constraint, with an analysis including higher-order modes potentially yielding more stringent limits.

\begin{figure}[htbp]
    \centering
    \includegraphics[width=1.0\columnwidth]{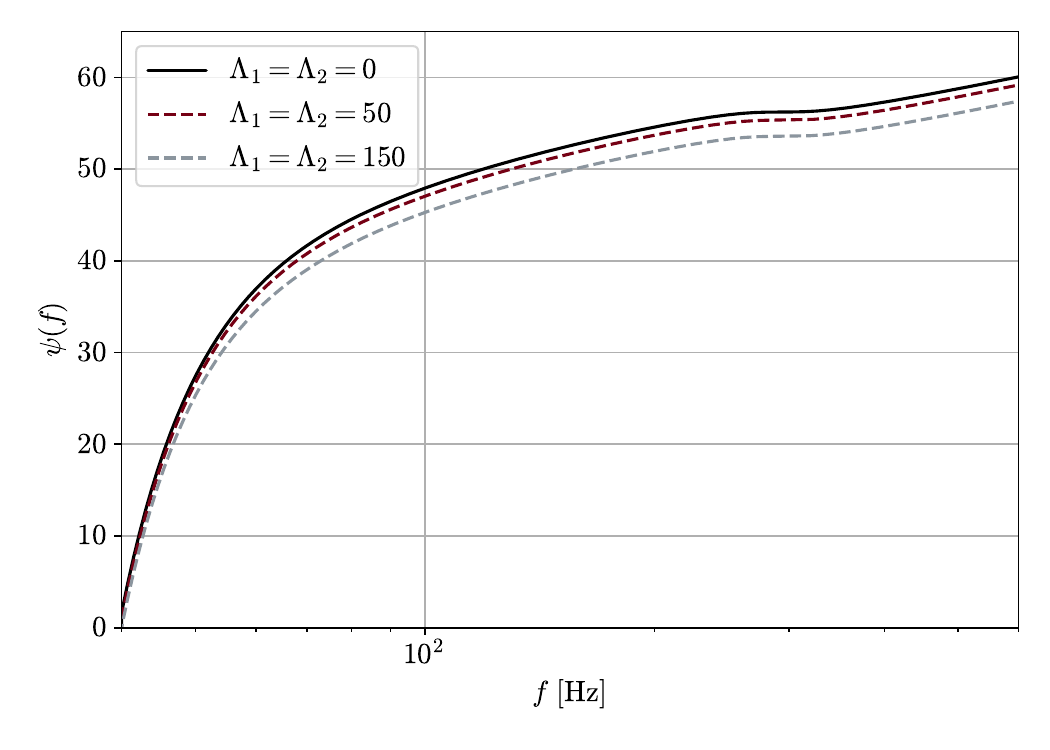}
    \caption{Comparison of the phase of the waveform for a non-spinning binary BH system with $m_1= 33.6~\msun$ and $m_2=32.2~\msun$ for the vacuum GR case and for the one augmented with TLNs for $\Lambda_1=\Lambda_2=50$ (red) and $\Lambda_1=\Lambda_2=150$ (gray).}
    \label{fig:Comaprison_WF}
\end{figure}

An example of a waveform's phase with the effect of non-zero TLNs in a system with  $m_1= 33.6~\msun$ and $m_2=32.2~\msun$, corresponding to the best source-frame component masses estimations for GW250114, is shown in Fig.~\ref{fig:Comaprison_WF}.

With this information we can now quantify the difference between the standard, zero TLN, waveform ($h_{\Lambda=0}$) and the one augmented with the non-vanishing tidal effects ($h_{\Lambda\neq0}$). This is typically done by computing the faithfulness, which is defined as 
\begin{equation}
    \mathcal{F}(h_{\Lambda=0},h_{\Lambda\neq0}) = \max_{t_{\rm ref},~ \phi_{\rm ref}}\frac{\langle h_{\Lambda=0}|h_{\Lambda\neq0}\rangle}{\sqrt{\langle h_{\Lambda=0}|h_{\Lambda=0}\rangle\langle h_{\Lambda\neq0}|h_{\Lambda\neq0}\rangle}}\, .
\end{equation}

\begin{figure}[htbp]
    \centering
    \includegraphics[width=\columnwidth]{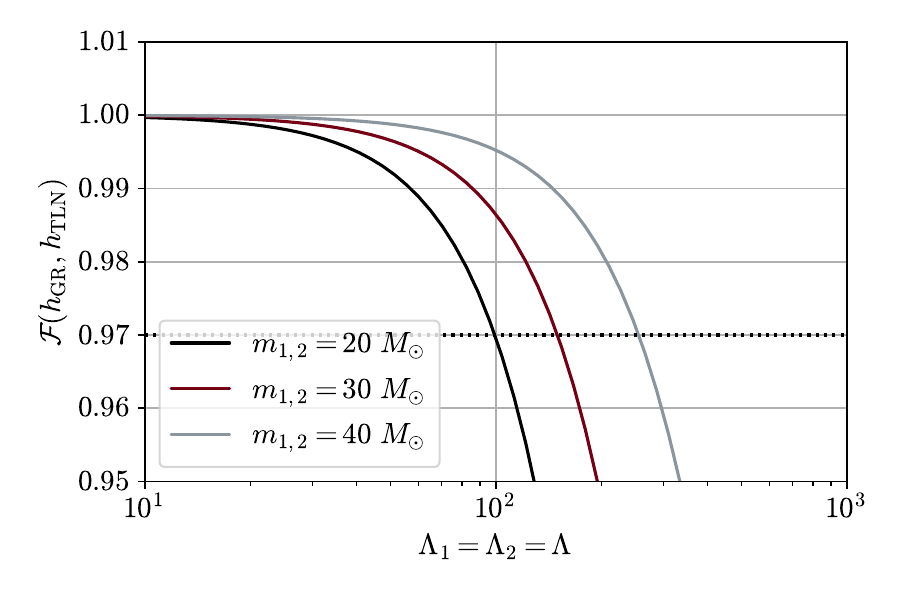}
    \caption{Faithfulness for equal-mass binaries for different tidal deformabilities.}
    \label{fig:faithfulness}
\end{figure}

This quantity varies from 0 to 1 and quantifies how similar two waveforms are. A general rule is that two waveforms are indistinguishable if their faithfulness is greater than 0.97. Figure~\ref{fig:faithfulness} shows the faithfulness for equal-mass binaries for different values of the tidal deformabilities and with the same PSD of the noise as for GW250114. It follows from it that lighter binaries show larger deviations from the GR vacuum case for a given value of $\Lambda$. Therefore, to set stringent upper limits, one would like a light binary, but with sufficient SNR to observe many cycles of the inspiral.

To perform the parameter estimation we employ a Bayesian analysis framework which relates the posterior distribution of some relevant parameters $\theta$ given the data $d$ and hypothesis $\mathcal{H}_i$, $p(\theta|d,\mathcal{H}_i)$, to the likelihood, $\mathcal{L}(\theta|d,\mathcal{H}_i)$, the prior $\pi(\theta|\mathcal{H}_i)$ and the Bayesian evidence $p(d|\mathcal{H}_i)$ as 
\begin{equation}
    p(\theta|d,\mathcal{H}_i) = \frac{\mathcal{L}(\theta|d,\mathcal{H}_i)\pi(\theta|\mathcal{H}_i)}{p(d|\mathcal{H}_i)}\, .
\end{equation}

We employ the standard likelihood under Gaussian noise as stated in Refs.~\cite{Ashton:2018jfp,vanderSluys:2008qx,vanderSluys:2007st,Veitch:2008ur}. The Bayesian evidence, 
\begin{equation}
    p(d|\mathcal{H}_i) = \int \pi(\theta|\mathcal{H}_i)\mathcal{L}(\theta|d,\mathcal{H}_i)~\td\theta \equiv \mathcal{Z}_i\, ,
\end{equation}
measures the marginal likelihood over all the prior distribution and quantifies the degree with which a model is capable of explaining the data under the prior and likelihood.The evidence allows us to compute the Bayes factor,  $\ln \mathcal{B}^{i}_{j} = \ln \mathcal{Z}_i-\ln\mathcal{Z}_j$, which can be used to select among two competing hypothesis, $\mathcal{H}_i$ and $\mathcal{H}_j$. 

The priors are reported in Tab.~\ref{tab:priors} for completeness and they represent the same priors used also in Ref.~\cite{KAGRA:2025oiz}. We also include the effect of the calibration uncertainties as stated in Refs.~\cite{Ashton:2018jfp,CalUncer}.

\begin{table*}[t]
\centering
\caption{Priors used in the parameter estimation. Angles are in radians.}
\label{tab:priors}
\renewcommand{\arraystretch}{1.15}
\begin{tabular}{llll}
\hline
\hline
\textbf{Parameter} & \textbf{Symbol} & \textbf{Prior} & \textbf{Support / Range} \\
\hline
Chirp mass  & $\mathcal{M}_c$ & UniformInComponentsChirpMass & $[29.94,\,32.08]~M_\odot$ \\
Mass ratio & $q=m_2/m_1$ & UniformInComponentsMassRatio & $[1/6,\,1]$ \\
Primary spin magnitude & $a_1$ & Uniform & $[0,\,0.99]$ \\
Secondary spin magnitude & $a_2$ & Uniform & $[0,\,0.99]$ \\
Primary spin tilt & $\theta_1$ & Sine$^{\dagger}$ & $[0,\,\pi]$ \\
Secondary spin tilt & $\theta_2$ & Sine$^{\dagger}$ & $[0,\,\pi]$ \\
Spin azimuthal diff. & $\phi_{12}$ & Uniform & $[0,\,2\pi)$ \\
JL-plane azimuth & $\phi_{JL}$ & Uniform & $[0,\,2\pi)$ \\
Luminosity distance & $d_L$ & UniformSourceFrame & $[10,\,10000]~\mathrm{Mpc}$ \\
Declination & $\delta$ & Cosine$^{\ddagger}$ & $[-\pi/2,\,\pi/2]$ \\
Right ascension & $\alpha$ & Uniform & $[0,\,2\pi)$ \\
Inclination & $\theta_{JN}$ & Sine$^{\dagger}$ & $[0,\,\pi]$ \\
Polarization & $\psi$ & Uniform & $[0,\,\pi)$ \\
Coalescence phase & $\phi_c$ & Uniform & $[0,\,2\pi)$ \\
Geocenter time & $t_c$ & Uniform & $[t_\mathrm{trig}-0.1,\,t_\mathrm{trig}+0.1]~\mathrm{s}$ \\
Primary tidal deformability & $\Lambda_1$ & LogUniform & $[10^{-3},5000]$ \\
Secondary tidal deformability & $\Lambda_2$ & LogUniform & $[10^{-3},5000]$\\
\hline
\hline
\end{tabular}

\vspace{4pt}
\raggedright
\footnotesize
$^{\dagger}$\ Sine prior: $p(\theta)\propto \sin\theta$ on $[0,\pi]$ (isotropic in direction). \\
$^{\ddagger}$\ Cosine prior: $p(\delta)\propto \cos\delta$ on $[-\pi/2,\pi/2]$ (isotropic on the sky). \\
\end{table*}

The parameter estimation is performed using the \texttt{Bilby} software with the \texttt{dynesty} nested sampler. In this way, not only we get samples from the posterior distribution, but we also get a good estimation of the Bayesian evidence that allows us to compute the Bayes factor reported.

\begin{figure*}[htbp]
    \centering
    \includegraphics[width=1.0\textwidth]{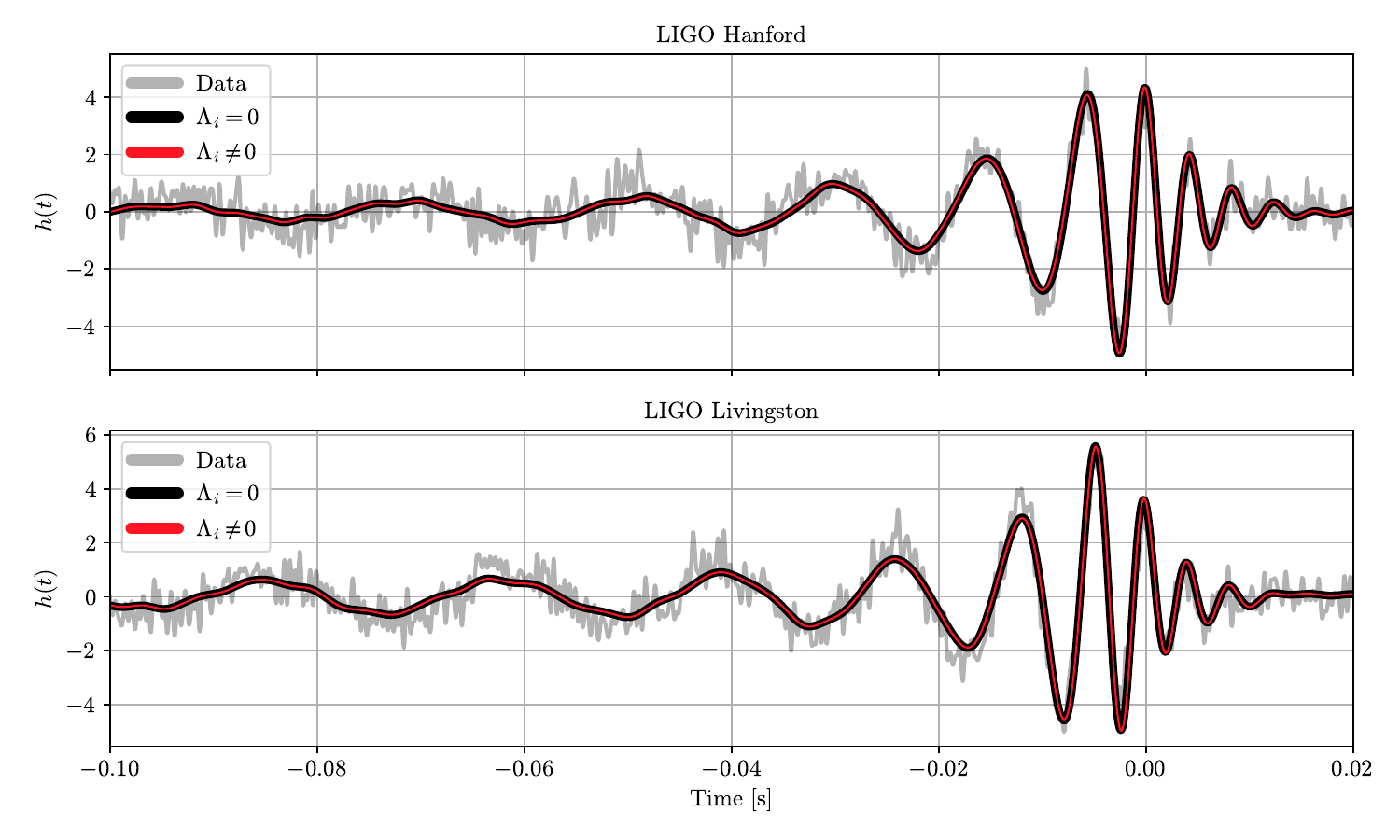}
    \caption{Comparison of the maximum log-likelihoodwaveform without TLNs ($\Lambda_i=0$) and the one containing the TLN deviations ($\Lambda_i\neq0$), compared to the real data for the two LIGO detectors.}
    \label{fig:Waveform_comparison_fit}
\end{figure*}

Similarly, we can compare the maximum likelihood waveform with the data both including and excluding the TLNs. This is displayed in Fig.~\ref{fig:Waveform_comparison_fit} and shows how the two waveforms are barely indistinguishable and both match the data. Both of them have a maximum likelihood $\ln \mathcal{L} = 2993.2$. As pointed out by the faithfulness even if there is some non-vanishing TLNs, for small values of the tidal deformability, the waveforms become indistinguishable.

\section{Priors on tidal deformabilities}
\label{app:priors}
In the main text we adopt a logarithmically uniform prior on the component tidal deformabilities,

\begin{equation}
\Lambda_i \sim \mathrm{LogUniform},[10^{-3},5000]\, ,
\end{equation}

which corresponds to a prior density

\begin{equation}
p(\Lambda_i) = \frac{1}{\Lambda_i , \ln(\Lambda_{\rm max}/\Lambda_{\rm min})}\, ,
\qquad
10^{-3} \le \Lambda_i \le 5000\, .
\end{equation}

This prior assigns equal probability mass to each decade in $\Lambda_i$, placing substantial prior support near the BH limit $\Lambda_i = 0$.
Although $\Lambda_i = 0$ lies formally outside the sampling range, the density diverges as $\Lambda_i \rightarrow 0$, ensuring that the prior does not artificially down-weights the region favored by the data for a BBH signal.
The LogUniform prior is particularly suitable for null-hypothesis tests, as it avoids imposing strong prior density at large $\Lambda_i$, which are disfavored by the likelihood and previous analyses. It also mitigates the known degeneracy between the effective tidal parameter $\tilde{\Lambda}$ and the mass ratio during the inspiral, making it effectively uninformative for order-of-magnitude constraints on $\Lambda_i$.
For completeness, we repeat the analysis using the more conventional linear prior, also used in Ref.~\cite{Chia:2023tle}:

\begin{equation}
\Lambda_i \sim \mathrm{Uniform},[0,5000],
\end{equation}
which corresponds to a density
\begin{equation}
p(\Lambda_i) = \frac{1}{5000},
\qquad
0 \le \Lambda_i \le 5000.
\end{equation}

\begin{figure*}[htbp]
\centering
\includegraphics[width=0.5\textwidth]{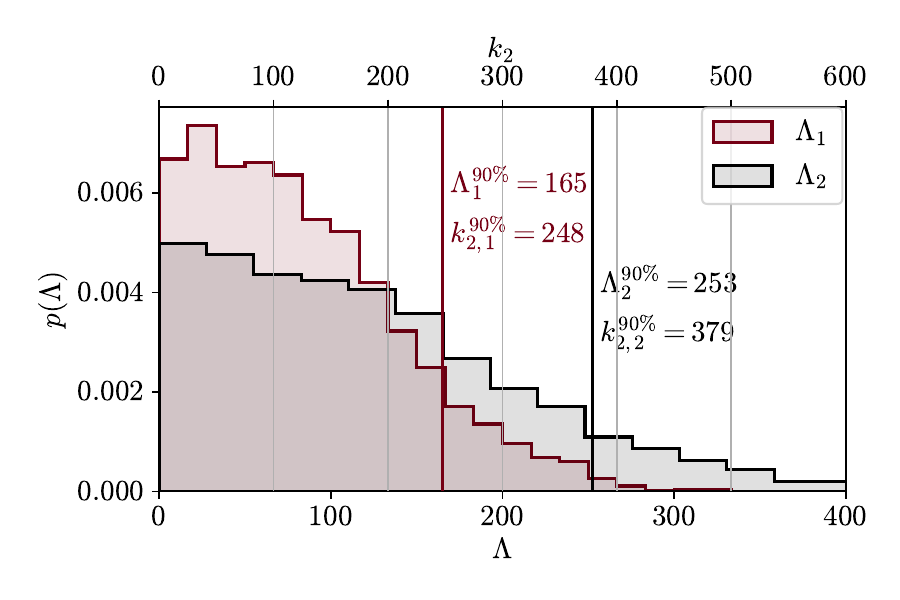}
\includegraphics[width=0.44\textwidth]{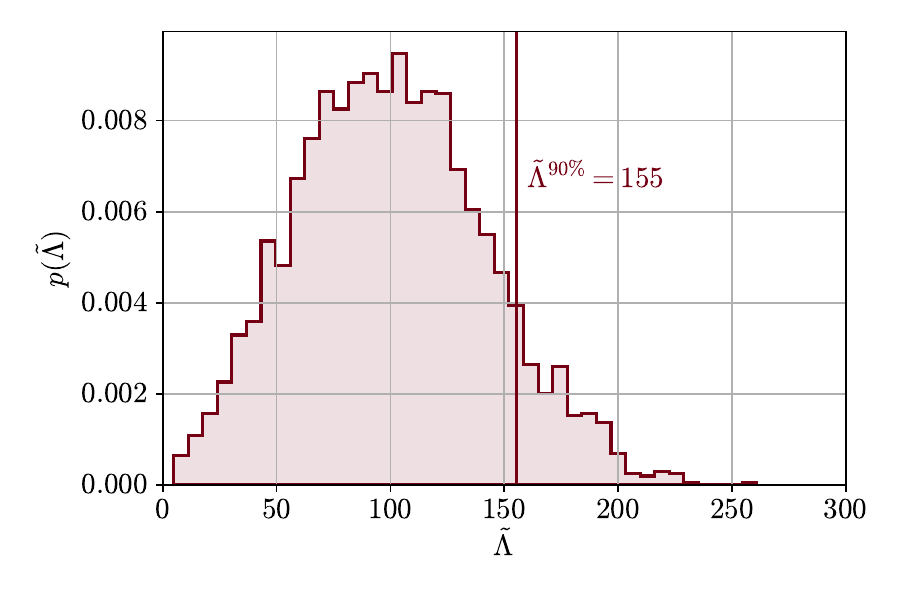}
\caption{\textit{(Left}) Marginalized posterior distributions for individual tidal deformabilities $\Lambda_i$ and tidal Love numbers $k_2$ for GW250114 using Uniform priors. Vertical lines indicate 90\% credible upper limits. (\textit{Right}) Marginalized posterior distribution for the effective tidal deformability $\tilde{\Lambda}$ using Uniform priors. The vertical line shows the 90\% credible upper bound.}
\label{fig:LambdaLimits_Uniform}
\end{figure*}

The Uniform prior spreads probability evenly over an extremely large region of parameter space in which the likelihood for a BBH signal is essentially zero. Consequently, the posterior is strongly affected by prior-volume effects, producing an apparent preference for finite tidal deformability, even though the data contain no such evidence. This behavior is evident in Fig.~\ref{fig:LambdaLimits_Uniform}: the posteriors are broader than in the LogUniform case, and the corresponding 90\% credible bounds are weaker:

\begin{align*}
\Lambda_1 & < 165, & k_{2,1} & < 248, \\
\Lambda_2 & < 253, & k_{2,2} & < 379, \\
\tilde{\Lambda} & < 155.
\end{align*}

In contrast, adopting the LogUniform prior (Fig.~\ref{fig:LambdaLimits} in the main text) yields posteriors that closely follow the likelihood, producing tighter upper limits and a more pronounced preference for vanishing tidal effects.

The different shapes of the $\tilde{\Lambda}$ posterior under the two priors can be intuitively understood. The LogUniform prior places substantial probability density at very small deformabilities, so the posterior naturally follows the likelihood, which peaks near $\tilde{\Lambda} \simeq 0$ for a BBH signal. This produces the characteristic “railing” against zero, fully reflecting the data-driven preference for vanishing tidal effects.
By contrast, the Uniform prior assigns a negligible prior density to the region near $\tilde{\Lambda}=0$ relative to the large volume at large deformabilities. Because the likelihood is essentially flat over most of the prior range, the posterior is dominated by prior-volume effects. This produces a broad distribution that peaks away from zero due to the statistical suppression of small $\tilde{\Lambda}$ values under the Uniform prior.
This prior-likelihood mismatch is reflected in the Bayesian evidence:

\begin{align*}
\ln \mathcal{B}^{\Lambda_i \neq 0}_{\Lambda_i = 0}  (\text{Uniform}) = -5.53 \pm 0.36, \\
\ln \mathcal{B}^{\Lambda_i \neq 0}_{\Lambda_i = 0}  (\text{LogUniform}) = -0.063 \pm 0.18.
\end{align*}

The Uniform prior strongly, yet misleadingly, favors the vanishing Love-number hypothesis, whereas the LogUniform prior provides a more conservative and data-driven assessment consistent with the null result.

\begin{figure}[htbp]
\centering
\includegraphics[width=\columnwidth]{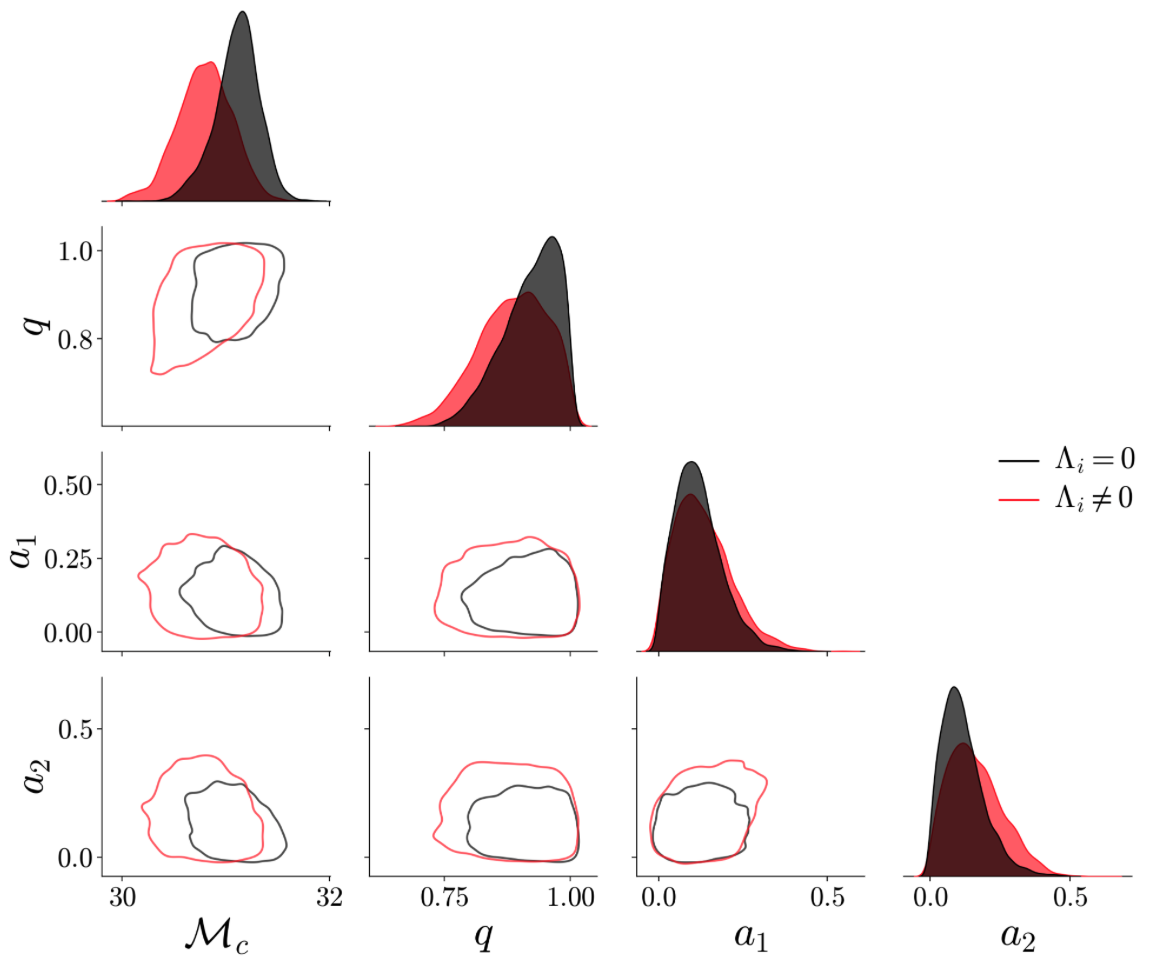}
\caption{Corner plot comparing the posterior distributions for the chirp mass, mass ratio, and spins with (red) and without (black) TLNs under a Uniform prior.}
\label{fig:corner_uniform}
\end{figure}

The impact of including tidal parameters on the inference of other source properties is illustrated in Fig.~\ref{fig:corner_uniform}, which compares the posterior distributions for the chirp mass $\mathcal{M}_c$, mass ratio $q$, and spins under the TLN (red) and non-TLN (black) hypotheses using a Uniform prior. The individual spin posteriors remain essentially unchanged, demonstrating that the spins are not affected by the choice of tidal model. In contrast, the posteriors for $\mathcal{M}_c$ and $q$ show a subtle but noticeable shift under the TLN hypothesis, because all TLN samples cannot be arbitrarily close to zero. 
This behavior contrasts with the LogUniform prior scenario (see main text), where the posterior support near $\Lambda_i \simeq 0$ allows the masses and spins to remain essentially unaffected by the inclusion of tidal parameters. 

Overall, this comparison highlights that the strong preference for the BBH hypothesis under the Uniform prior is dominated by prior-volume effects, not by information in the data. For these reasons, the LogUniform results are reported in the main text as the more appropriate choice for null-hypothesis tests.

\begin{figure}[htbp]
\centering
\includegraphics[width=1.0\columnwidth]{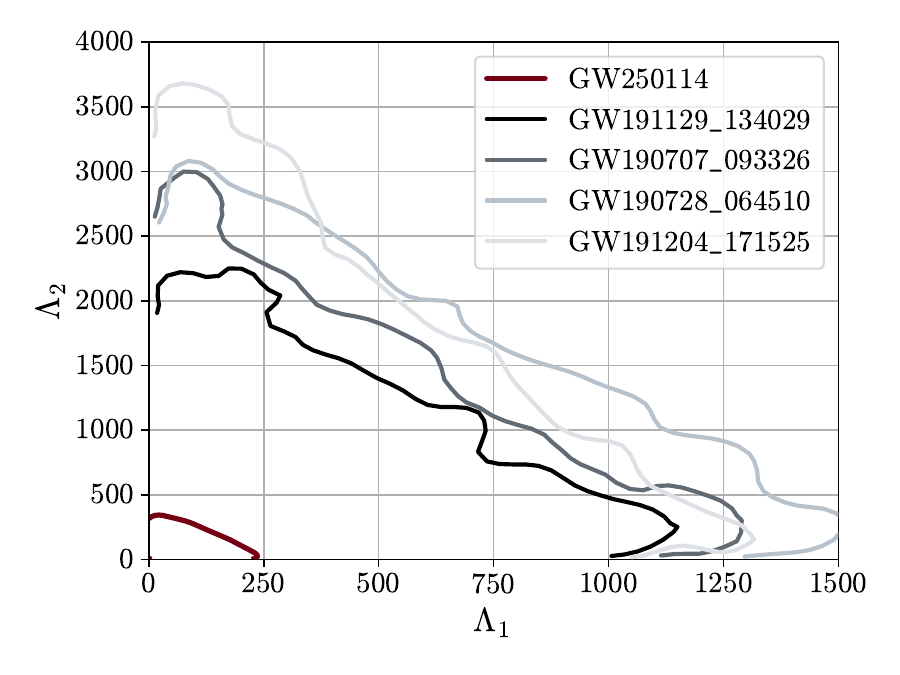}
\caption{Joint 90\% credible-level constraints on component tidal deformabilities compared with O3 results from Ref.~\cite{Chia:2023tle}. The lower-left corner corresponds to the vanishing-TLN scenario expected for BHs.}
\label{fig:LambdaComparison}
\end{figure}

Finally, Fig.~\ref{fig:LambdaComparison} compares our constraints on the component tidal deformabilities with those obtained from O3 BBH events.
Our results, shown from the Uniform prior analysis for consistency with the results in Ref.~\cite{Chia:2023tle}, remain fully compatible with $\Lambda_i = 0$ and still represent the most stringent simultaneous limits to date, with support concentrated in the lower-left corner of the $(\Lambda_1, \Lambda_2)$ plane corresponding to the BBH scenario. 

\section{Injection results}

To validate our analysis and assess potential biases introduced by neglecting tidal effects, we perform an injection study using Gaussian noise with the same PSD as GW250114. We inject signals with component tidal deformabilities $\Lambda_1=\Lambda_2 \in \{ 0, 200, 1000\}$, and recover the signals using both a non-tidal ($\Lambda_i=0$) and a tidal ($\Lambda_i\neq0$) model. The injected parameters correspond to the mode of the posteriors obtained with the IMRPhenomPv2 without TLNs and we only recover the masses, spins and tidal deformabilities, fixing the rest to the injected values during inference. Priors for these recovery parameters are identical to those used for the real event and Uniform in tidal deformability components.

\begin{figure}[htbp]
    \centering
    \includegraphics[width=1.0\columnwidth]{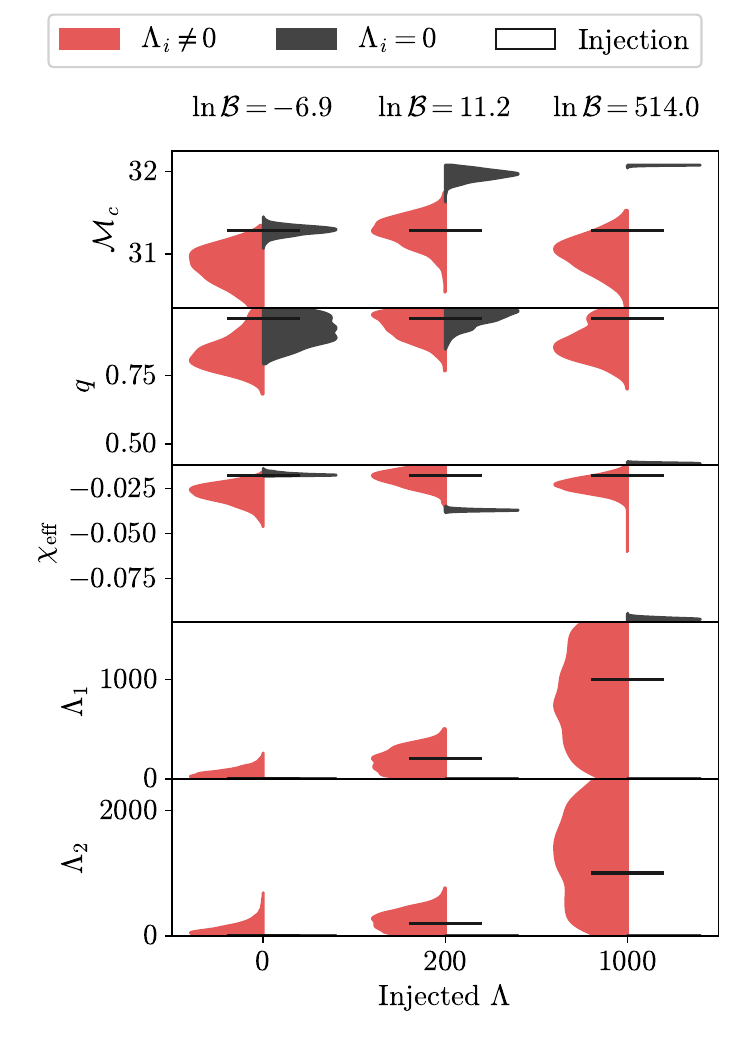}
    \caption{Violin plots comparing the posterior distributions of key parameters ($\mathcal{M}_c$, $q$, $\chi{\rm eff}$, $\Lambda_i$) for injections with $\Lambda_1=\Lambda_2 \in \{ 0, 200, 1000\}$. The horizontal lines indicate the injected values.}
    \label{fig:violin_injections}
\end{figure}

Figure~\ref{fig:violin_injections} summarizes the recovered posteriors for the injections. For the zero-TLN injection, recovery with a non-tidal model correctly recovers all parameters, while recovery with a tidal model introduces biases: the chirp mass $\mathcal{M}_c$ and mass ratio $q$ are shifted to lower values due to the degeneracy with $\tilde{\Lambda}$, which propagates to a smaller effective spin $\chi{\rm eff}$, with the posterior largely inconsistent with the injected values. For the $\Lambda_1=\Lambda_2=200$ injection, the non-tidal recovery overestimates $\mathcal{M}_c$ and underestimates $\chi{\rm eff}$, while $q$ is recovered adequately. The tidal recovery reproduces all parameters correctly. For the high-TLN injection $\Lambda_1=\Lambda_2=1000$, the non-tidal recovery exhibits more extreme biases, including a downward shift in $q$, whereas the tidal recovery captures the injected values but with broader posteriors and slight bimodality in $q$, reflecting the increasing challenge of recovering large tidal deformabilities.
These results demonstrate that neglecting tidal effects can systematically bias mass and spin parameters, particularly at high $\Lambda_i$, whereas including the tidal model correctly accounts for the correlations and recovers the injected values within uncertainties.

\begin{figure}[htbp]
    \centering
    \includegraphics[width=1.0\columnwidth]{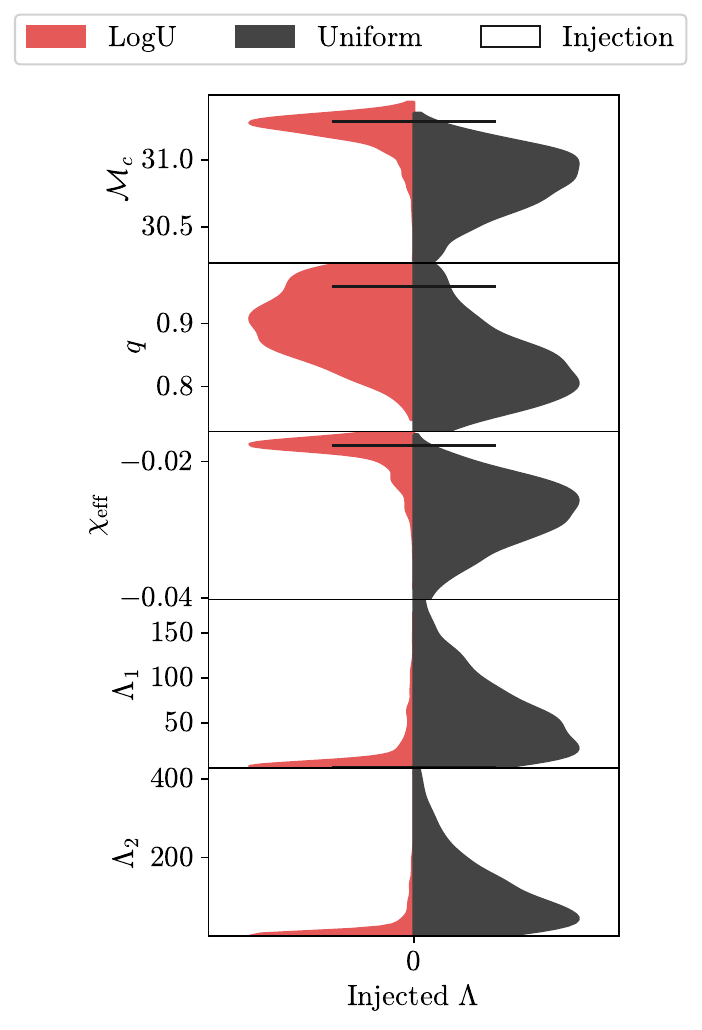}
    \caption{Violin plot comparing the result of an injection with vanishing TLNs with a LogUniform[$10^{-3},5000$] prior and with a Uniform$[0,5000]$ prior.}
    \label{fig:injections_prior}
\end{figure}

Figure~\ref{fig:injections_prior} compares the posteriors of key parameters obtained from two analyses of a simulated signal with vanishing TLNs, using tidal models but different priors on the deformabilities. The LogUniform[$10^{-3},5000$] prior accurately recovers the injected values: the posteriors for $\Lambda_i$ “rail” against zero, and the mass and spin parameters are unbiased. In contrast, the Uniform[$0,5000$] prior spreads probability over a larger prior volume at high $\Lambda_i$, producing posteriors for the deformabilities that peak away from zero and inducing small but noticeable positive biases in the recovered masses, illustrating the prior-driven effects discussed for the real-event analysis.

\section{Environmental model}
We employ an analogous setup and parameterization to that of Refs.~\cite{DeLuca:2022xlz,DeLuca:2024uju,DeLuca:2025bph}, in which it is assumed that both BHs are surrounded by an environment. In such works, the effective tidal deformability on an equal mass binary with its components immersed in the same type of environment, is modeled as 
\begin{equation}    \tilde{\Lambda}=\frac{2}{3}\mathcal{F}\varepsilon\tilde{L}^5\, ,
\end{equation} 
where the term $\varepsilon$ measures the fraction of mass of the environment compared to that of the BH, $\tilde{L}$ its characteristic length normalized by the BH mass and $\mathcal{F}$ encodes the nature of the environment. For example, as discussed in Ref.~\cite{DeLuca:2025bph}, $\mathcal{F}= 1$ would correspond to a thin shell of pressureless dust surrounding a Schwarzschild BH, while $\mathcal{F}=15/2$ would correspond to a gravitational atom in the lowest energy mode. Since GW250114 is close to be a symmetric binary system, we model it in the same way to be consistent. This parametrization is fully agnostic in the sense that only uses simple scalings relations and keeps the particularities encoded in the prefactor, allowing to rescale the results as appropriate.

As noted in the main text, treating the TLNs as constant across the entire inspiral is a simplifying assumption for the case where an environment is present, as it neglects the tidal disruption expected once the components reach sufficiently small separations. An environment that dresses the BHs would also increase the BH masses, and those augmented masses would themselves be subject to disruption as well. We deliberately omit these effects because our analysis is agnostic to any specific reinterpretation. To minimize bias, we therefore restrict our conclusions to the region where the environment’s disruption frequency lies above the \texttt{IMRPhenomPv2} transition between the inspiral and intermediate phases.

For an equal-mass binary the cut frequency can be estimated to be 
\begin{equation}
    f_{\rm cut} = \sqrt{\frac{2}{\gamma^3}}\frac{1}{\pi~ m_{\rm BH}}\sqrt{\frac{1+\varepsilon}{\tilde{L}^3}}\approx\frac{2.4\times10^4~\mathrm{Hz}}{m_{\rm BH}/M_\odot}\tilde{L}^{-3/2}\, ,
\end{equation}
where $\gamma = 2.44$ for fluid bodies, and where we have assumed that $\varepsilon\ll 1$ so it is negligible. Evaluating this for $\tilde{L}=6$ and for the average mass one gets roughly $f_{\rm cut}\sim 50 $~Hz. At frequencies beyond this one, tidal disruption would effectively vanish the tidal deformability. Additionally, there is a higher limit that one could theoretically probe on $\tilde{L}$, which corresponds to that in which the tidal disruption happens at a frequency smaller than those being probed. According to Ref.~\cite{DeLuca:2025bph} this is 
\begin{equation}
    \tilde{L}\leq\frac{1}{\gamma}\left[ \frac{\sqrt{2}}{\pi ~m_{\rm BH}~f_{\min}} \right]^{2/3}\, ,
\end{equation}
which for $f_{\min}=20$~Hz and the average mass of GW250114, implies $\tilde{L}\lesssim 11$. Accessing larger environments would require reaching lower frequencies, making next-generation ground- and space-based experiments great at probing them~\cite{DeLuca:2025bph,ET:2025xjr}.

The inspiral phase of the \texttt{IMRPhenomPv2} waveform that we use ends at $f=0.018c^3/(GM)\sim 50$~Hz~\cite{Khan:2015jqa}. This implies that around $\tilde{L}\sim6$, not accounting for tidal disruption might still be acceptable, but as $\tilde{L}$ increases, $f_{\rm cut}$ becomes smaller than the transition frequency of the waveform and tidal disruption might become relevant. Therefore, while the accessible range is $6\lesssim\tilde{L}\lesssim11$, only the lower end of this range can be trusted. In Ref.~\cite{DeLuca:2025bph}, the systematic biases of using a waveform that does not account for the changes in the environment during the inspiral has been studied and can produce errors in the estimated parameters at the $\mathcal{O}(10\%)$ level, being $\tilde{\Lambda}$ the parameter most affected.

\section{Boson star models}
Boson stars are complex bosonic configurations that are bound together by the effect of gravity. The simplest models are governed by an action of the form 
\begin{equation}
    S=\int\td^4x \sqrt{-g}\left[ \frac{R}{16\pi}-g^{ab}\partial_a\Phi^*\partial_b\Phi-V(|\Phi|^2)\right]\, .
\end{equation}

Following the models described in Ref.~\cite{Cardoso:2017cfl}, we consider three different scalar potentials that enter the action, which lead to three types of BSs: the minimal BS~\cite{Kaup:1968zz}, the massive BS~\cite{Colpi:1986ye} and the solitonic BS~\cite{Friedberg:1986tq}. The potential for the minimal BS is a simple $V(|\Phi|^2)=\mu^2|\Phi|^2$, where $\mu$ represents the mass of the boson. The massive BS has a potential of the form $V(|\Phi|^2)=\mu^2|\Phi|^2+\frac{\alpha}{4}|\Phi|^4$, where $\alpha$ denotes the quartic self-coupling. Finally, the soliton potential reads as $V(|\Phi|^2) = \mu^2|\Phi|^2\left[ 1-\frac{2|\Phi|^2}{\sigma_0^2}\right]^2$ where $\sigma_0$ is a coupling parameter.

In Ref.~\cite{Cardoso:2017cfl} the electric quadrupolar and octupolar TLNs are reported alongside the magnetic ones, but we focus only on the electric quadrupolar TLN, as this is the one that is tested in our analysis. The same reference provides the tabulated data of the TLNs for various values of the product of $m\mu$, as reported in Ref.~\cite{DataBS}. The range of $m\mu$ provided in such files covers the stable branch of the possible solutions.

In order to set the upper limits reported in the main text, we linearly interpolate the value of $m\mu$ that matches with our 90\% confidence level upper limit of the electric TLN, $k_{2,i}$. In the case of the minimal BS, the most compact configuration, $m\mu\simeq0.633$, produces an electric TLN of $k_2\simeq 113$, which is larger than our reported upper limit, which allows to discard the BS nature of this object under the minimal coupling. In Ref.~\cite{Cardoso:2017cfl}, it was already predicted that LIGO would be able to completely rule out minimal BSs mergers, and our results confirm it with an actual event. 

 For a massive BS with $\alpha=100$ ($\alpha=10^3$, $\alpha=10^4$) the most stable configuration is $m\mu\simeq 0.696$ ($m\mu\simeq1.144$, $m\mu =3.124$)~\cite{Cardoso:2017cfl} which correspond to $k_2=68.3$ ($k_2=25.8$, $k_2=13.6$). Finally, the solitonic BS has a last stable configuration at $m\mu = 3.125$ corresponding to a $k_2=53.8$. These values already rule out the BS–BS interpretation for the massive case with $\alpha=100$ and for the solitonic model with $\sigma_0=0.05$, as the primary component is excluded at 90\% credibility.

We derive constraints on the fundamental boson mass, $\mu$, for the massive cases with $\alpha=10^3$ and $\alpha = 10^4$ by inverting the theoretical relation for each posterior sample of mass and TLN from our analysis. Any samples with $k_2$ values outside the model's physical range are discarded. The resulting credible intervals on $\mu$ are therefore conditional on the object being a BS of that specific type.

These results are only valid for non-rotating BSs, as spin could slightly modify the TLNs. Because GW250114 exhibits low effective spin, this approximation is expected to be adequate.

\end{document}